\documentclass[a4paper,twocolumn]{esapub2005} % European paper
\usepackage{float}
\usepackage{amsmath}
\usepackage{amssymb}
\usepackage{times}
\usepackage{graphicx}
\usepackage[numbers]{natbib}
\setcitestyle{square}
\usepackage{url}
\usepackage{soul}
\usepackage{xcolor}
\usepackage{hyperref}

\title{Autonomous Robotic Arm Manipulation for Planetary Missions using Causal Machine Learning}
\author{Cian McDonnell}
\affil{MSc Astronautics and Space Engineering, Cranfield University, United Kingdom, cian@mcdonnell.eu}
\author{Miguel Arana-Catania}
\author{Saurabh Upadhyay}
\affil{School of Aerospace, Transport and Manufacturing, Cranfield University, United Kingdom,  miguel.aranacatania@cranfield.ac.uk, saurabh.upadhyay@cranfield.ac.uk}

\begin{document}


\maketitle

\begin{abstract}

Autonomous robotic arm manipulators have the potential to make planetary exploration and in-situ resource utilization missions more time efficient and productive, as the manipulator can handle the objects itself and perform goal-specific actions. We train a manipulator to autonomously study objects of which it has no prior knowledge, such as planetary rocks. This is achieved using causal machine learning in a simulated planetary environment. Here, the manipulator interacts with objects, and classifies them based on differing causal factors. These are parameters, such as mass or friction coefficient, that causally determine the outcomes of its interactions. Through reinforcement learning, the manipulator learns to interact in ways that reveal the underlying causal factors. We show that this method works even without any prior knowledge of the objects, or any previously-collected training data. We carry out the training in planetary exploration conditions, with realistic manipulator models.
\end{abstract}

\section{Introduction}
\label{section:introduction}
Autonomous manipulation has significant potential in planetary missions, as it can increase the amount of time spent exploring the environment and doing science activities. We review current methods for autonomous manipulation of objects, both on Earth and in planetary exploration environments. We see that approaches to manipulation of unknown objects usually must leverage a large amount of training data to work.

Reinforcement learning is commonly used to teach a robot manipulator certain skills, such as pick and place operations on objects, to solve specific tasks. In \cite{SkillAcquisition:1} the manipulator chooses its policy from
a set of actions, and over time learns to use more of those actions that gave the best
results in the past. These are scored based on how close the object gets to a target position. Here we follow a different causal approach \cite{Sontakke:1} using reinforcement learning to find which actions give the most information about the ``causal factors", the main parameters that determine the dynamics of the objects, so that this knowledge can be used to carry out any general task. The manipulator learns which actions produce the most distinguishable interactions for each factor. For example, it may learn about an object's frictional properties by pushing it along the ground, and studying the distance travelled, which is directly affected by the friction.

There are relatively few papers on autonomous manipulators in planetary environments. Typically, two types of objects are considered in the existing literature: known objects where the robot has prior knowledge (e.g. size, shape, mass of a known scientific device), and objects which are completely unknown to the robot, e.g. planetary rocks. In \cite{LRU_One:1}, the authors present a design for a light-weight rover that can pick up and assemble known objects. The objects are detected based on machine-learning classification by colour. The same rover is used in \cite{LRU_Two:1} to demonstrate the placement, testing and collection of payload instruments, in a simulated lunar environment on Earth.
\\

More sophisticated techniques are required to learn to manipulate unknown objects. In \cite{Lunar_Grasping:1}, the authors train a robot to grasp objects on the Moon, using 3D octree representations of the environment. A convolutional neural network is used in conjunction with actor-critic reinforcement learning to allow the robot to detect objects, choose the most efficient pose, and pick the objects up. The policy, learned in simulation, is then passed to a real-world rover in a lunar analogue centre. 

Finally, it is also helpful to look at work on the problem of manipulation in terrestrial environments. While this is not carried out with planetary exploration in mind, many techniques can be transferred, especially when the work involves manipulation of unknown objects. In \cite{Scalable_Grasping:1}, several real-world robots are used to attempt to grasp unknown objects, using a single colour camera as input. The grasp attempts are then passed to a self-supervised reinforcement learning algorithm, producing a policy that allows grasping with high success rate.

Our analysis of the previous up-to-date research on autonomous planetary exploration with manipulators concludes that the major challenges in the area are related to the following categories: Prior Knowledge of Objects, Generalisability to Planetary Environments, and Applicability to Non-Grasping Operations

\textbf{Prior Knowledge of Objects.} In \cite{LRU_One:1} and \cite{LRU_Two:1}, the rover described is designed only to manipulate previously known objects (the base station parts, and payload instruments), using colour and expected dimensions to classify them. These approaches are not applicable to unknown object operations, given the lack of previous knowledge. The approaches used in \cite{SkillAcquisition:1}, \cite{Lunar_Grasping:1} and \cite{Scalable_Grasping:1} require comparatively less prior knowledge.

\textbf{Generalisability to Planetary Environments.}
The model described in \cite{Scalable_Grasping:1} draws on training data from many real-life grasping operations, carried out on Earth.  This is impractical for planetary missions, as any training data generated on Earth would be very different from the data produced in extraterrestrial environments and thus bias the training of the machine learning models. In \cite{SkillAcquisition:1}, a similar approach is used, but the training is carried out in a virtual environment, causing the same problems with generalisability. \cite{Lunar_Grasping:1} also trains the robot in a simulation and then applies the model to real life (a ``sim-to-real" transfer). However, this strategy only has a success rate of about 32\% in the best case. This is likely due to the simulation data not perfectly matching the data in the real world.  For \cite{LRU_One:1} and \cite{LRU_Two:1}, as reinforcement learning is not used, the method can in principle be applied to an extraterrestrial environment as well as a terrestrial one.

\textbf{Applicability to Non-Grasping Operations.} These works mainly focus on grasping the objects. While \cite{LRU_One:1} shows that the rover can also carry out assembly operations, other complex manipulations, such as rolling, or rotating the object, are not explored. The same occurs in \cite{LRU_Two:1}. \cite{Lunar_Grasping:1} and \cite{SkillAcquisition:1} each focus entirely on grasping. It is mentioned in the discussion of \cite{Scalable_Grasping:1} that the method can be generalised to other operations, but the topic is not discussed further.

In Table \ref{tab:existing_approaches} we present the summary of the previous comparative study of the most relevant needs for the goal of autonomous robot manipulation in planetary environments.

\begin{table}[!ht]
\caption{Comparison of existing approaches based on three identified requirements.}
    \begin{tabular}{|p{3.05cm}|p{1.25cm}|p{1.0cm}|p{1.25cm}|}
    \hline
      \textbf{Existing Work} & \multicolumn{3}{l|}{\textbf{Key Requirements}}\\
      \hline
        & Unknown Objects? & Genera-lises? & Non-Grasping? \\
      \hline
      Liu et al. \cite{SkillAcquisition:1} &Yes  &  & \\ \hline
       Schuster et al. \cite{LRU_One:1} &  & Yes & \\ \hline
      Lehner et al.  \cite{LRU_Two:1} &  & Yes & \\ \hline
    Orsula et al. \cite{Lunar_Grasping:1} & Yes & &\\ \hline
      Kalashnikov et al. \cite{Scalable_Grasping:1} & Yes & & Yes\\ \hline
    \end{tabular}
        
    \label{tab:existing_approaches}
\end{table}

As shown in Table \ref{tab:existing_approaches}, the existing approaches do not fulfil all three key requirements at once. In particular, approaches that can manipulate unknown objects are, overall, not generalisable to planetary environments. This is due to the need to leverage data that has been previously collected in a very different environment.

As the major contribution of this work, we investigate an interaction-based causal learning approach, described in \cite{Sontakke:1} as a prospective candidate to fulfil all three of the requirements at the same time. We define the relevant problem with a realistic manipulator and environment constraints. By trying many different actions to measure a specific property of an object, and focusing only on those that give the most information about the object, a manipulator can autonomously learn about its surroundings in any type of environment. This is possible without needing any prior training data, or prior information about the objects themselves. Furthermore, because the manipulator is not explicitly told which actions it should try, it can discover or apply the learning to alternative manipulations to grasping. From here we carry out simulations to demonstrate that the investigation approach can be applied to the identification of different parameters. 

The remainder of the paper is as follows: Section 2 describes the problem, with the manipulator model and environment details given. Section 3 discusses the use of causal machine learning to solve the relevant problem. The test cases are simulated in Section 4 and concluding remarks are presented in Section 5.

\section{PROBLEM DESCRIPTION}
\label{section:problemdescription}
\subsection{Determination of Causal Factors}
A causal factor is a parameter of an environment, that causally affects the result of a particular action of the manipulator on the environment. They are parameters such that, by applying a certain sequence of actions on the environment, the observations obtained are organised in distinguishable disjoint sets according to the parameter values. For example, if the manipulator pushes an object, the outcome will vary depending on the mass of the object - a heavy object will be harder to push than a lighter one, and will not be pushed as far. Here the causal factor is the mass.

In particular, if the \textit{same} action is repeated in several different environments, it should in principle be possible to measure or classify the underlying causal factor associated with each environment, according to the observed result of such an action. This will be the problem to be solved in this work.

Being able to determine causal factors fulfils all three requirements discussed in Section \ref{section:introduction}. The manipulator does not \textit{a priori} need to know any of the physical properties of the objects with which it interacts to learn about them (requirement 1), no prior training data of the environment is required making it fully generalisable (requirement 2), and it carries out operations other than grasping (requirement 3).

\subsection{Environment Simulation and Constraints}
A simulation in PyBullet\footnote{~\url{https://pybullet.org}} was used to implement the determination of causal factors. The simulation design included manipulator design and constraints as well as environmental parameters adequate to represent a real-life planetary exploration scenario.

\textbf{Manipulator Model.} The robot manipulator used in the simulations was designed to be analogous to the arm of NASA's Curiosity rover. This design was used as Curiosity's arm was made to carry out functions similar to those in this work, such as scooping surface samples \cite{Anderson:1}. 

\begin{figure}[!ht]
    \centering
    \includegraphics[width=0.8\linewidth]{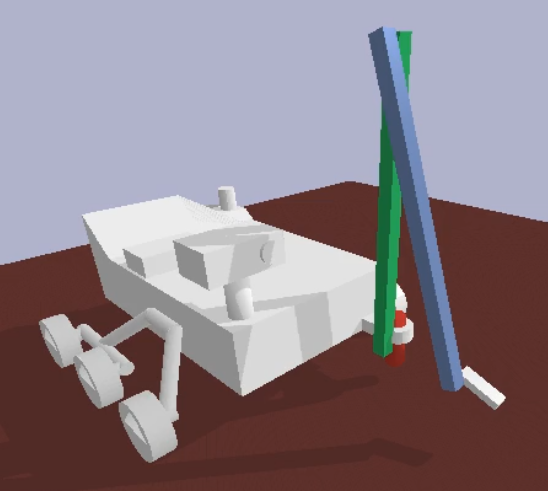}
    \caption{Manipulator used in simulations.}
    \label{fig:manipulator}
\end{figure}
\begin{figure}[!ht]
\centering
\includegraphics[width=0.8\linewidth]{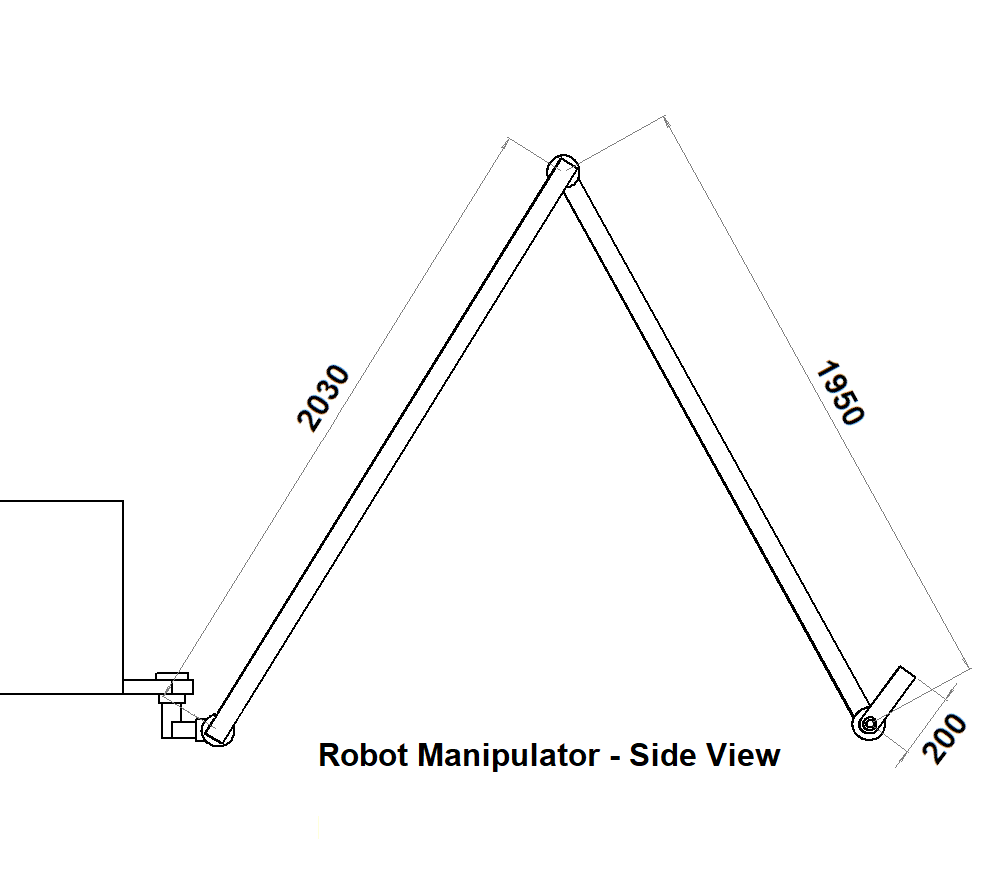}
\caption{CAD Drawing of Manipulator, showing dimensions of links (in mm).}
\label{fig:manipulator_cad}
\end{figure}

As seen in Figure \ref{fig:manipulator}, the manipulator can control the azimuthal angle or ``compass direction" of its end effector, using the red azimuth actuator. It can also control the height and radial distance from the end effector to the rover, using the green and blue links shown. A ``shoulder" joint connects the green link to the red azimuth actuator, while an ``elbow" joint connects the green and blue links.
A CAD drawing of the manipulator is shown to illustrate the dimensions of these links, in Figure \ref{fig:manipulator_cad}.

\textbf{Joint angle limits.} Joint angle limits were required for the manipulator to work, ensuring that the links would not end up in unrealistic or unphysical configurations. The angle at the ``shoulder" between the red and green links was locked to the range [0°, 150°], and the angle at the ``elbow" between the green and blue links was locked to the range [-180°, 0°], so that the blue link always pointed downwards.

\textbf{Joint torques.} Limits on the torque that could be exerted by the joints were also necessary, as applying control algorithms with default torque limits generally leads to very unrealistic results. The maximum torque exerted by a motor is directly related to the power provided to it, so these gave an estimate of what the torque should be. 

The power output of a motor is given by
\begin{equation}
    \label{equation:motor_power}
    P = \tau\omega
\end{equation}
where $P$ is the power output, $\omega$ is the angular velocity, and $\tau$ is the torque exerted. Power subsystem data from the Perseverance rover \cite{Perseverance:1} was used to estimate the maximum possible torque that can be generated. Using the dimensions of the manipulator as in Figure \ref{fig:manipulator_cad}, and a maximum end-effector velocity of 2 m s$^{-1}$, we find a maximum torque of 460 N m for the azimuth actuator and shoulder joint, and 240 N m for the elbow joint.

\section{Proposed methodology: Causal Machine Learning}
Here we present the implementation of our proposed solution to the problem defined in Section \ref{section:problemdescription} following the approach in \cite{Sontakke:1}.

\subsection{Interaction-Based Learning Overview}
To implement the interaction-based learning and allow the classification of objects based on different causal factors, the program required three broad steps:
\begin{enumerate}

    \item Simulate a set of actions, each in several random environments that represent different classes of values of the causal factors (E.g. heavy and light mass). Record the position time series of the object in each simulation.
    \item For each action, apply machine learning time series clustering algorithms to classify the corresponding set of time series into distinct clusters or classes. Score the actions based on the accuracy of the classification of their time series.
    \item Using the scores in step 2, choose a number of best or ``elite" actions. Plan new actions that are similar to the elite actions. Repeat step 1 using the new actions, until the maximum number of repetitions/iterations or required accuracy in the classification is reached.
\end{enumerate}
These steps teach the manipulator the most informative actions to use to learn about its environment (measured by the ability of the action to classify by causal factor of interest). By choosing only the actions that give the most accurate classifications, with each iteration the manipulator improves its ability to learn. After iterating many times, the code returns the best actions produced. This method is similar to that seen in \cite{SkillAcquisition:1}, which also chooses actions similar to those that performed well in the past.

\begin{figure}[!ht]
\centering
\includegraphics[width=1\linewidth]{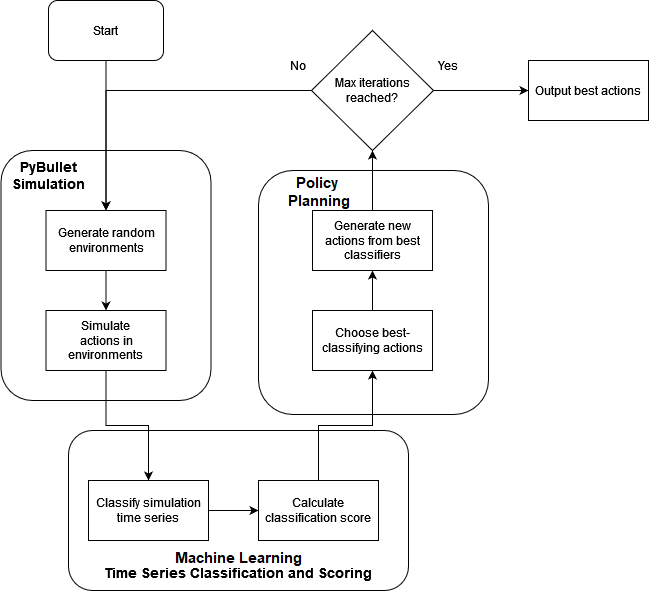}
\caption{Flowchart showing different algorithm steps.}
\label{fig:algorithm_flowchart}
\end{figure}
Figure \ref{fig:algorithm_flowchart} shows the relationships between the three steps described above. These will be discussed in the sections that follow.

\subsection{Physics Simulation}
The package used to study the manipulator was PyBullet. The simulator automatically models physical phenomena such as collisions, friction and gravity, and at the same time, it provides full customizability of each of these phenomena. 

\begin{figure}[!ht]
\centering
\includegraphics[width=0.8\linewidth]{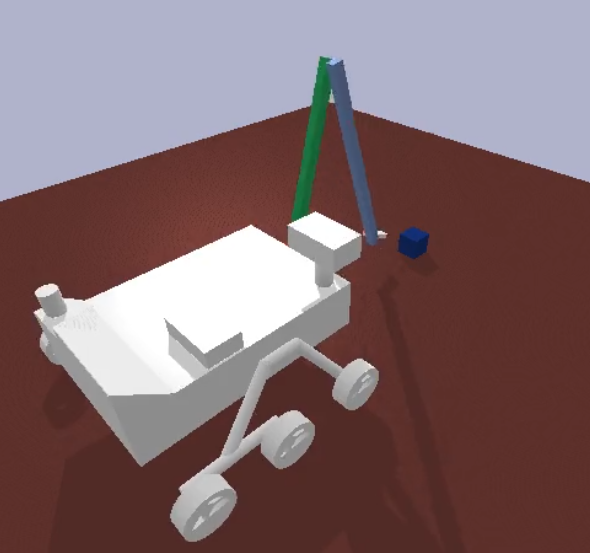}
\caption{Starting setup used in Test Case 1.}
\label{fig:starting_setup}
\end{figure}

\textbf{Environment setup.} In each simulation, three meshes were loaded in PyBullet - the robot manipulator, the object being studied (a cube), and the ground plane, which was static. The positions and orientations of each were the same every time. The cube's parameter under study was chosen from a random distribution representing the ranges of interest to analyse. It was chosen from a bimodal distribution, consisting of two Gaussian peaks, representing two main categories for the value. The means and standard deviations of these peaks were chosen so that there was not too much overlap between them (the difference between the means was three standard deviations). This ensured that the environments could in principle be separated into two distinct classes or clusters. Once the environment was set up, the manipulator was allowed to execute its action. The starting setup is shown in Figure \ref{fig:starting_setup}.

\textbf{Definition of Actions.} Each possible action of the manipulator was represented as a list of vectors in $\mathbb{R}^3$, with each vector representing a change in position. Beginning from the initial position of the manipulator's end effector, the action is executed by changing the end effector's position by the first vector, then by the second, then the third, etc., until the end of the list is reached. This is achieved using PyBullet's built-in inverse kinematics calculator. Because the vectors in each action can have real number entries, they can vary continuously, and so this definition gives a relatively large space of actions. In the first iteration of the simulation, there are no previous actions to choose from, so the actions were chosen randomly - choosing each entry of the action's vectors from a uniform distribution. This represents the initial lack of information about which actions are relevant to classify the parameter.

\textbf{Simulation Output.} The output of the simulation was the full position vector time series of the object as the manipulator acted upon it. This was used in the next step of the algorithm - the time series classification and scoring.

\subsection{Time Series Classification and Scoring}
After the action had been executed on a set of objects, each with a different causal factor, the set of time series of the objects was separated into two classes using machine learning time series k-means clustering using the tslearn library implementation\footnote{~\url{https://tslearn.readthedocs.io}}. The metric used was the Euclidean distance. The clustering identified two distinct groups in the set of objects, and assigned a label to each object based on which group it belonged to. If the two groups are well-separated, then the action is effective at separating the objects based on causal factors. For example, if an action causes light objects to be separated from heavy objects by time series k-means clustering, it is effective in determining the mass of the objects.

\begin{figure}[!ht]
\centering
\includegraphics[width=1\linewidth]{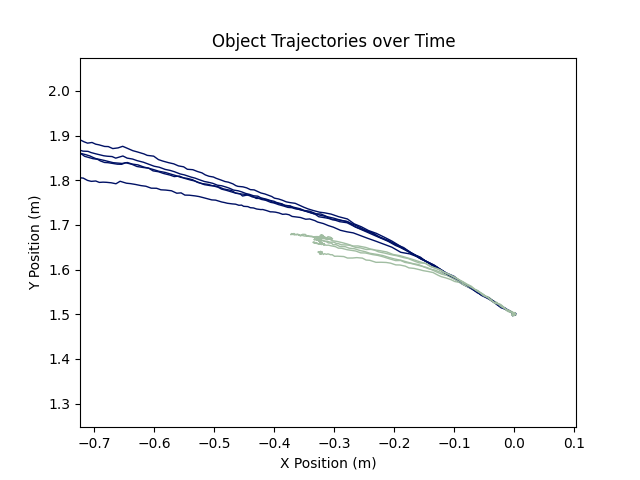}
\caption{Time series of objects labelled by k-means clustering.}
\label{fig:causal_factors}
\end{figure}
Figure \ref{fig:causal_factors} shows a set of object time series, labelled by k-means clustering. The heavier objects, coloured green, do not move as far when pushed by the manipulator. The lighter blue objects move much further. This difference in behaviour can be detected by the clustering algorithm and hence used to classify the objects.

\textbf{Reward function.} The approach implements a reward function used to rank which actions give the best results for the task at hand.
In this work, the reward function involved two terms. The first was the F1 score \cite{FScore:1}, a metric which compares the labels assigned by k-means clustering to the true groups to which the objects belong. It takes values between 0 and 1, with higher values meaning a better identification of each object in its true class according to the causal factor.

The second term was the silhouette score \cite{silhouette:1}, a measure of how well-separated the two clusters were. A silhouette score of 1 implies perfect separation, while a silhouette score of 0 implies no separation. This was used to measure how robust is the separation between the classes and to penalise random classifications.

The relative weight of the F1 score and silhouette score in the final reward function was determined by Equation \ref{equation:rewardfunction}:
\begin{equation}
\label{equation:rewardfunction}
    R = (1-\alpha)F + \alpha S
\end{equation} 
where $R$ is the overall reward, $F$ is the F1 score, $S$ is the silhouette score, and $\alpha$ is the reward function parameter. A value of 0.3 for $\alpha$ was used. This ensures the reward function mainly depends on the F1 score, while giving enough value to the silhouette score to remove random classifications.

\subsection{Policy Planning}
To find the best actions, the cross-entropy method was employed. The method is a Monte Carlo technique used to choose an element of a set that maximises a reward function. The approach works by generating in each iteration random actions in order to explore all the possible actions of the manipulator, but focusing in each iteration on generating similar actions to the ones that obtained the highest reward in the previous iteration step.

An implementation of this method to optimize a function is described as Algorithm 4.1 in \cite{crossentropy:1}. The function is evaluated at several random samples, drawn from a multidimensional Gaussian distribution with mean $\boldsymbol{\hat{\mu}}_{t-1}$ and standard deviation $\boldsymbol{\hat{\sigma}}_{t-1}$ (t-1 referring to the index of this iteration). Then, the samples with the highest value of the function (``elite samples", denoted by the $e$ subscript) are averaged to a mean sample $\boldsymbol{\hat{\mu}}_e$. The standard deviation $\boldsymbol{\hat{\sigma}}_e$ is also taken. Finally, the samples for the next iteration are again drawn from a multidimensional Gaussian distribution. The mean $\boldsymbol{\hat{\mu}}_t$ and standard deviation $\boldsymbol{\hat{\sigma}}_t$ of this distribution are smoothed between the initial parameters and the elite parameters, using a smoothing parameter $\beta$:
\begin{align}
\label{equation:crossentropy}
    \boldsymbol{\hat{\mu}}_t &=  (1-\beta)\boldsymbol{\hat{\mu}}_{t-1}  + \beta \boldsymbol{\hat{\mu}}_e\\
    \boldsymbol{\hat{\sigma}}_t &= (1-\beta)\boldsymbol{\hat{\sigma}}_{t-1}  + \beta \boldsymbol{\hat{\sigma}}_e \nonumber
\end{align}

Over many iterations, the overall set of samples drawn in each iteration converges to a sample that locally maximises the relevant function, and the standard deviation decreases as this happens. 

Along with the machine learning time-series classification and reward function, this method is crucial to allow convergence to the best-classifying actions. By generating actions that are similar to the best actions of the previous iteration, we improve the probability of finding good actions in the next iteration, and can quickly narrow down the search space to only the actions that are likely to give useful results.

\section{Simulation Results}
Given the algorithm described in the previous sections, the following three test cases are considered. The simulations were carried out on a PC with the following specifications: AMD Ryzen 5 3600X CPU, EVGA GeForce GTX 1060 6GB GPU, and 16GB Aegis G.SKILL DDR4 RAM.

\subsection{Test Case 1: Mass}
In the first test case, the mass of the object was varied. A cube of side length 25 cm was used. The ``light" and ``heavy" clusters had mean masses of 25 kg and 50 kg respectively, giving realistic densities for planetary rocks \cite{density:1}. The performance of the algorithm was studied while varying the number of actions used per iteration of the algorithm.

\begin{figure}[!ht]    
    \centering
    \includegraphics[width=\linewidth]{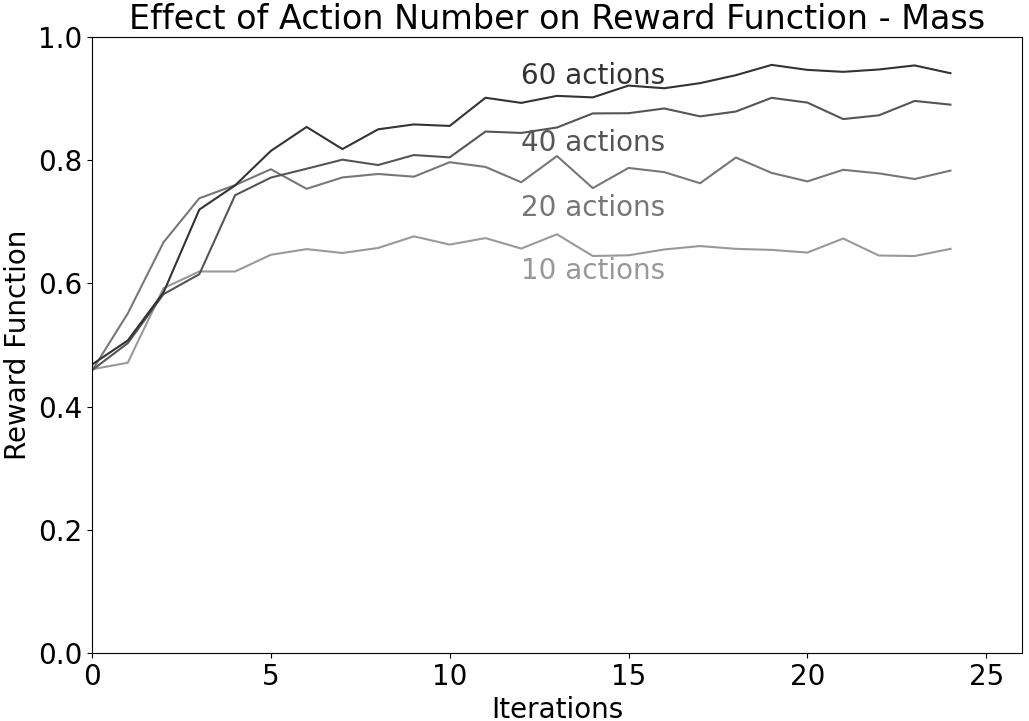}
    \caption
    {Evolution of reward functions in Test Case 1.}
    \label{fig:test_case_1_action_comparison}
\end{figure}

The reward function over time for different numbers of actions, averaged over 16 simulations, is shown in Figure \ref{fig:test_case_1_action_comparison}. Table \ref{tab:test_case_1_action_results} shows the average final reward $R$, F1 score $F$, and silhouette score $S$ at the end of each simulation.

\begin{table}[!ht]
\centering
\caption{Final scores from Test Case 1.}
    \begin{tabular}{|p{2cm}|p{1cm}|p{1cm}|p{1cm}|}

    \hline

      Actions&\textbf{$R$} & \textbf{$F$} & \textbf{$S$}\\ \hline

     10 & 0.656 & 0.775 & 0.379\\ \hline
     20 & 0.783 & 0.867 & 0.587\\ \hline
      40 & 0.890 & 0.938 & 0.779\\ \hline
          60 & 0.941 & 0.973 & 0.868\\ \hline
    \end{tabular}
        
    \label{tab:test_case_1_action_results}
\end{table}

From Figure \ref{fig:test_case_1_action_comparison}, we can see that the number of actions used hugely affects the algorithm's performance. With only 10 actions, the performance is rather poor, reaching an average reward function of only 0.656. This implies that many of the simulations do not converge to an action that correctly classifies the objects. However, by the nature of the cross-entropy method used to plan new actions, the algorithm will always converge to \textit{some} action. Hence, it is essential to find a good classifier in the early stages of the algorithm. If not, it can become ``stuck" on a poorer action that locally maximises the reward, despite not classifying properly.

It is for this reason that using more actions improves the performance. At 60 actions, the average reward improves to 0.941. With more actions, we can more thoroughly explore the space of possible actions to use, increasing the probability of finding good classifiers.

\subsection{Test Case 2: Friction}

In the second test case, the friction coefficient between the object and the ground was varied. The same cube as before was used, this time with a fixed mass of 50 kg. There were two possible clusters of friction coefficients, with means of 0.4 and 0.6, both realistic values for planetary rocks \cite{friction:1}.

As before, the performance of the algorithm was studied while varying the number of actions used per iteration. The actions used to differentiate the clusters were also observed to record differences to those used in Test Case 1.

\begin{figure}[!ht]
    \centering
    \includegraphics[width = \linewidth]{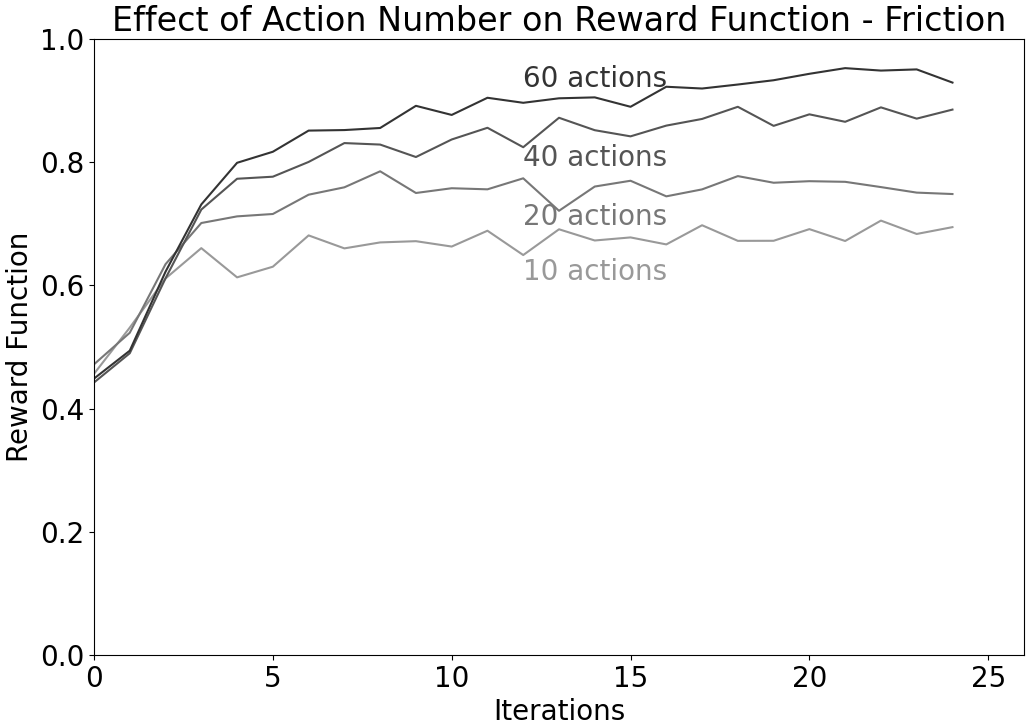}
    \caption{Evolution of reward functions in Test Case 2.}
    \label{fig:test_case_2_comparison}
\end{figure}

The reward function over time  for different numbers of actions, averaged over 16 simulations, is shown in Figure \ref{fig:test_case_2_comparison}. Table \ref{tab:test_case_2_results} shows the average final reward $R$, F1 score $F$, and silhouette score $S$ at the end of each simulation.

\begin{table}[!ht]
\centering
 \caption{Final scores from Test Case 2.}
    \begin{tabular}{|p{2cm}|p{1cm}|p{1cm}|p{1cm}|}

    \hline

      Actions&\textbf{$R$} & \textbf{$F$} & \textbf{$S$}\\ \hline

     10 & 0.695 & 0.811 & 0.423\\ \hline
     20 & 0.748 & 0.842 & 0.530\\ \hline
     40 & 0.885 & 0.937 & 0.765\\ \hline
     60 & 0.929 & 0.968 & 0.839\\ \hline

    \end{tabular}
       
    \label{tab:test_case_2_results}
    
\end{table}

As with Test Case 1, the evolution of the reward function improves as the number of actions used is increased. Despite the change in the causal factor, the results are very similar between the two test cases.

After analysing the optimal actions for classification found, we observe that there is a change in the type of actions found between the first and second tests. In Test Case 1, there were many actions that sufficed to classify the objects based on mass. The object could be pushed, rolled, or even flipped into the air. When the causal factor is friction, however, the best classifiers found tend to be gentle pushes along the ground. This is because the friction coefficient only affects the outcome of actions when the object is in contact with the ground. Hence, the best actions will be those that maximise the time the object is grounded.
    
\subsection{Test Case 3: Gravity}
For the final test case, the object's parameters were kept fixed, while the acceleration due to gravity, $g$, for the whole environment was varied. The clusters of values for $g$ were chosen from the mean lunar and Martian values of 1.625 m s$^{-2}$ \cite{MoonGravity:1} and 3.721 m s$^{-2}$ \cite{MarsGravity:1}, with slight variations ($\sigma$ = 0.01 m s$^{-2}$) due to the topography of these bodies. Again, the number of actions per iteration was varied and the nature of the actions used to differentiate the causal factors was studied.

\begin{figure}[!ht]
    \centering
    \includegraphics[width = \linewidth]{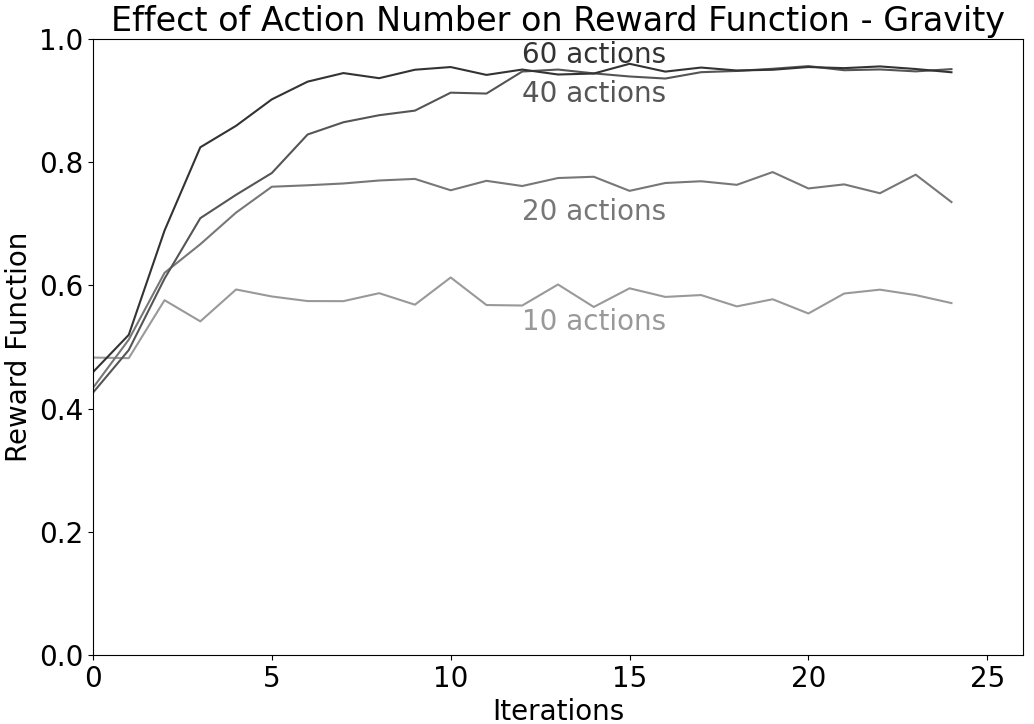}
        \caption{Evolution of reward functions in Test Case 3.}

    \label{fig:test_case_3_comparison}
\end{figure}

The averaged reward function over time for different numbers of actions is shown in Figure \ref{fig:test_case_3_comparison}. Table \ref{tab:test_case_3_results} shows the average final reward $R$, F1 score $F$, and silhouette score $S$ at the end of each simulation.

\begin{table}[!ht]
\centering
     \caption{Final scores from Test Case 3.}
    \begin{tabular}{|p{2cm}|p{1cm}|p{1cm}|p{1cm}|}

    \hline

      Actions&\textbf{$R$} & \textbf{$F$} & \textbf{$S$}\\ \hline

     10 & 0.571 & 0.705 & 0.261\\ \hline
     20 & 0.735 & 0.805 & 0.572\\ \hline
     40 & 0.951 & 0.969 & 0.910\\ \hline
     60 & 0.946 & 0.961 & 0.912\\ \hline

    \end{tabular}
   
    \label{tab:test_case_3_results}
    
\end{table}

The performance in Test Case 3 is worse than the previous test cases when fewer actions are used, e.g. 10 actions, but much better when 40 or 60 actions are used. In particular, the silhouette score is much better on average. This is likely due to the minimal variation between environments within the same class, e.g. two environments of lunar gravity. The difference in gravity is so small that environments of the same class behave almost identically, giving a perfect separation between the two classes and driving up the value of $S$. The actions discovered by the manipulator to differentiate the environments tended to be gentle pushes along the ground, as in Test Case 2. These are sufficient, as the normal force between the object and ground (and hence the friction) is directly affected by the acceleration due to gravity.
%\subsection{Optional: worst case, computation complexity,}

\section{Discussion}

The results of the simulations verify the requirements set out in Section \ref{section:introduction}. The algorithm works without the manipulator needing access to any information on the object's physical parameters. It can thus handle completely unknown objects, satisfying the first requirement. Regarding the second requirement, no previous training data is needed for the reinforcement learning to work, only the information generated through its own simulations. Hence the method can generalise to planetary environments. Finally, the manipulator has full autonomy over the actions it uses. Due to this, it does not need to grasp objects to learn about them, fulfilling the third requirement mentioned.

The algorithm in this project could be used in other types of interactions, such as for example the ones related to wheel slip prediction - the measurement and prediction of the slipping of rover wheels on planetary soils. In particular, there is a significant amount of work in the area of ``in advance" wheel slip estimation \cite{Basri:1}, which surveys oncoming terrain for potential hazards that could cause the rover to become stuck. Here the causal factors would be the mechanical properties of the soil. The manipulator could be pushed along the ground, and the forces measured to estimate these properties. The technique would allow the classification of the terrain into ``Safe" or ``Unsafe" classes, without the rover needing to drive over it first.

Other approaches that attempt to learn about objects by training the manipulator to carry out one \textit{specific} action tend to have very sparse reward functions. That is, only very specific inputs give significant rewards. For example, an approach that trains the manipulator to grasp and then lift the object to measure the mass requires a very specific sequence of actions to work properly. Thus, hundreds of thousands of actions can be needed during reinforcement learning for these approaches, as seen in \cite{Lunar_Grasping:1} and \cite{Scalable_Grasping:1}, which both teach the manipulator to grasp.

The tests in this work show that, if the objective is to learn to manipulate an object, using causal factors can considerably shorten the process. An object's mass, the causal factor studied in Test Case 2, affects the outcome of almost any kind of interaction the manipulator can carry out - pushing, lifting, rolling, etc. Hence, if we wish to learn about an object's mass, any action that interacts with the object in some way is likely to give some information about it. This offers an advantage over previous methods in learning about the environment.

\section{Conclusions}
We have used a causal machine learning-based approach that allows a robot manipulator to learn autonomously about its surroundings in a simulated planetary environment. This method works by revealing the differences in interactions due to changing causal factors of the environment. Reinforcement learning is used to choose those actions that give the best performance in separating the environments. We have shown that this allows classification of objects based on changing parameters, and that it generalises to different types of parameters.

\section*{Acknowledgments}
The work is partially supported by MSc Astronautics and Space Engineering at Cranfield University, and by UKRI/EPSRC TAS-S: Trustworthy Autonomous Systems: Security node (EP/V026763/1).

% The following bibliography was produced with

%\bibliography{esapub}
% The results are inserted directly here to simplify
% the demonstration.

\bibliography{references}
\bibliographystyle{ieeetr}

\end{document}